\documentclass{IEEEtran}
\usepackage{cite}
\usepackage{amsmath,amssymb,amsfonts}
\usepackage{algorithmic}
\usepackage{graphicx}
\usepackage{textcomp}

\def\BibTeX{{\rm B\kern-.05em{\sc i\kern-.025em b}\kern-.08em
    T\kern-.1667em\lower.7ex\hbox{E}\kern-.125emX}}
\begin{document}
\title{Load Balanced Dynamic Resource Allocation for MTC Relay}
\author{Yifu Yang, \IEEEmembership{Student Member, IEEE}, Gang Wu, \IEEEmembership{Member, IEEE}, Weidang Lu, \IEEEmembership{Member, IEEE}, Yizhong Zhang, \IEEEmembership{Student Member, IEEE}
\thanks{(\emph{Corresponding author: Gang Wu.}) }
\thanks{Y. Yang, G. Wu and Y. Zhang are with the National Key Laboratory of Science and Technology on Communication, University of Electronic Science and Technology of China, Chengdu 611731, China (e-mail: yifuyang@std.uestc.edu.cn; wugang99@uestc.edu.cn; yizhong@std.uestc.edu.cn).}
\thanks{W. Lu is with the College of Information Engineering, Zhejiang University of Technology, Hangzhou 310058, China (e-mail: luweid@zjut.edu.cn).}
}

\maketitle

\begin{abstract}
A Load Balancing Relay Algorithm (LBRA) was proposed to solve the unfair spectrum resource allocation in the traditional mobile MTC relay. In order to obtain reasonable use of spectrum resources, and a balanced MTC devices (MTCDs) distribution, spectrum resources are dynamically allocated by MTCDs regrouped on the MTCD to MTC gateway link. Moreover, the system outage probability and transmission capacity are derived when using LBRA. The numerical results show that the proposed algorithm has better performance in transmission capacity and outage probability than the traditional method. LBRA had an increase in transmission capacity of about 0.7dB, and an improvement in outage probability of about 0.8dB with a high MTCD density.
\end{abstract}


\section{Introduction}
\label{section1}
\IEEEPARstart{I}{n} the machine-type communication (MTC) scenario, the radio access network (RAN) will be congested due to the large number of MTC devices (MTCDs) accessing the data aggregation center (DAC) simultaneously. Different approaches have been proposed to alleviate the problem, i.e. prioritized random access, access class barring and distributed queuing. Another potential solution is data aggregation\cite{b1,b2}, some MTCDs form a group and send data to the DAC through the MTC gateway (MTCG).

The performance of MTC relay has been extensively studied. The uplink average data rate of the MTC relay was studied under different spectrum allocation schemes, by use constraint gradient ascent optimization algorithms\cite{b3}. In \cite{b4}, an ALOHA protocol for multi-hop networks is proposed to reduce latency by optimizing the coverage of each relay. In \cite{b5}, stochastic geometry was used to study the effect of reducing system delay when different MTCG selection schemes were used. Moreover, resource waste can be reduced when using data bundling on MTCG\cite{b6}.

Resource allocation in MTC relay is necessary to improve system transmission capacity and reduce the outage probability. Spectrum efficiency can be improved, and more device connections can be supported by sharing spectrum resources in a non-orthogonal way within the group\cite{b2}. In \cite{b7}, a channel-aware resource scheduling was proposed, gateways tend to allocate resources to MTCDs with better channel state, which improves the transmission success rate. MTCG prioritizes MTCD according to different parameters (such as quality of service (QoS)), and allocates resources to MTCD based on priority, which can also improve the transmission capacity of the system\cite{b8}. In \cite{b9}, the trade-off between transmission capacity and fairness of resource allocation was studied, a global optimal resource allocation scheme is proposed to improve network throughput.

However, the above literatures didn't consider dynamic allocation resources on MTCD-MTCG (MTCD2G) link when analyzing system performance. Therefore, this paper studies the dynamic resource allocation scheme based on MTC relay, aiming to reduce the outage probability and increase the transmission capacity when supporting massive MTCDs connection. The main contributions of this paper are summarized as follows:

\begin{itemize}

\item	A load balancing relay algorithm (LBRA) was proposed. In this algorithm, MTCD is first grouped by random geometry method, and then the MTCD in each group is regrouped based on the load of groups.

\item The transmission capacity and outage probability of the system when using the proposed algorithm are derived, and compared with the simulation results.

\end{itemize}
%
%
%
%
%
%
%
%
%
\section{System Model}
\label{section2}
\begin{figure}[!t]
\centerline{\includegraphics[width=\columnwidth]{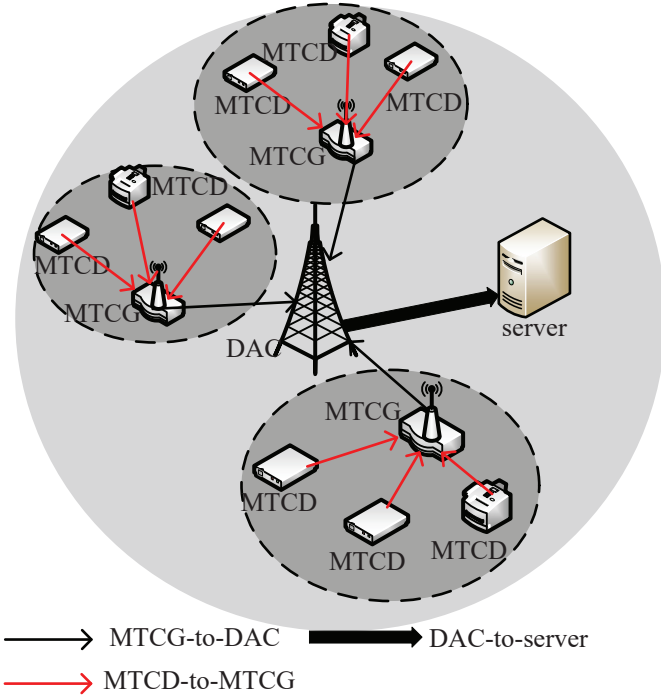}}
\caption{System model.}
\label{fig1}
\end{figure}

As shown in Fig. \ref{fig1}, a mobile MTC relay covered by a DAC is considered. MTCDs have the potential to become relays due to their excellent data processing and communication capabilities. All MTCDs and nearest MTCG form a group, following the DAC’s decision. The positions of MTCD and MTCG are assumed to obey two independent homogeneous poisson point process (HPPP), $\Phi_{D}=\left\{X_{i}\right\}, \Phi_{G}=\left\{Y_{i}\right\}$, and the distribution density are $\lambda_D$ and $\lambda_G$, respectively. This paper focuses on the rational scheduling of resources within MTCD groups to improve system performance.

MTC is allocated spectrum resources by DAC in resource block (RB). The RBs allocated by DAC for MTC are assumed to have \emph{R$_{m}$}, and are divided into \emph{R$_{1}$} and \emph{R$_{2}$} for MTCD2G link and MTCG-DAC (MTCG2C) link respectively.

For simplicity, all MTCDs in MTCD2G link are assumed to use the same modulation and coding scheme, and a single data packet of a fixed size is transmitted at power \emph{P}, the path loss is considered. Each data channel is composed of $\omega_{1}$ RBs, and these RBs are sufficient to send a data packet, then there are $U_{1}=\left\lfloor\frac{R_{1}}{\omega_{1}}\right\rfloor$ data channels in total. The path loss model is $l(r)=r^{-\alpha}$, where $\alpha$ is the path loss index and \emph{r} is the distance between the transmitter and receiver distance. MTCG can decode the packet successfully when the signal-to-interference ratio (SIR) of a MTCD signal is greater than the threshold $\eta$.

The number of MTCG is represented by \emph{G}, and the specific value of \emph{G} is known by the DAC. MTCG2C link is divided into \emph{G} data channels. With $\omega_{2}$ to indicate the RB required to send a single data packet to the DAC, then, at most $U_{2}\left(Y_i\right)=\left\lfloor\frac{\gamma_{i}\times R_{2}}{\omega_{2}}\right\rfloor$ data packets can be relayed by the MTCG at $Y_{i}$ to the DAC, $\gamma_{i}$ is the spectrum division coefficient, which is used to indicate the spectrum allocated by DAC to MTCG, it is proportional to the area of each grouping region and satisfy $\sum_{0}^{G-1} \gamma_{i}=1$. MTCG at $Y_{i}$ randomly selected $U_{2}(Y_{i})$ packets to be relayed to the DAC if the number of successfully decoded packets is greater than $U_{2}(Y_{i})$, otherwise, all data packets are relayed to the DAC.

With the outage probability of MTC relay is $\varepsilon$, using the definition of \emph{transmission capacity} in \cite{b10}
\begin{equation}
C=\lambda_{D}(1-\varepsilon)
\label{eq1}
\end{equation}
where $\lambda_{D}$ represents the distribution density of MTCD, the transmission capacity here does not consider the specific transmission rate of each link, but only consider whether each link can satisfy the quality of service requirements of the device.
%
%
%
%
%
%
%
%
\section{Load Balancing Relay Algorithm}
\label{section3}
\begin{figure}[!t]
\centerline{\includegraphics[width=\columnwidth]{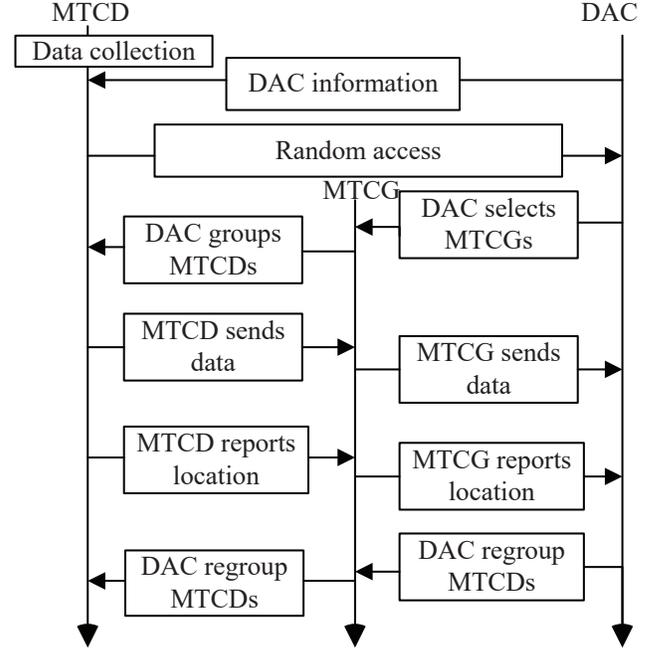}}
\caption{The flow chart of LBRA.}
\label{fig2}
\end{figure}

LBRA is proposed in this paper for the uneven distribution of MTCD. In the MTCD2G link, to make the number of MTCD in each group relatively balanced and MTCG2C link resources can be used reasonably, for the two nearest groups, MTCD will always transfer from the large group to the small group. The algorithm flow is shown in Fig. \ref{fig2}. MTCD completes random access and sends location information to DAC after collecting data, according to the information broadcast by DAC. MTCG is selected by the DAC to complete the grouping and allocate spectrum resources based on this information. Afterwards, the MTCD's location changed due to its mobility, and it needed to resend location information to the DAC for regrouping.
The data packets sent by MTCD can be successfully relayed to the DAC, when 1, 2, 4, 5 or 1, 3, 4, 5 occur simultaneously in the following five events, without loss of generality, MTCG is assumed at $Y_{0}\in\mathrm{\Phi}_G$.

\begin{enumerate}
\item The typical MTCD is the nearest MTCD (on the channel \emph{u}) to an MTCG at $Y_{0}$;
\item MTCD is transferred from the nearest $Y_{i}$ into the $Y_{0}$ area;
\item MTCD is transferred from the $Y_{0}$ area to the nearest $Y_{i}$ area;
\item Note that the MTCD sending data on channel \emph{u} after regrouping is $Z_{i}^{u}$, and the MTCG located at $Y_{0}$ successfully captures the packet;
\item Data packets captured by MTCG at $Y_{0}$ can be successfully relayed to DAC.
\end{enumerate}

Suppose a typical MTCD at $X_{i}$ and sends a packet on the channel $u \in\left\{1,2, \dots, U_{1}\right\}$. For simplicity, use $\mathcal{V}_{X_{i}^{u}, Y_{0}}^{u}$ to represent event 1, use $\mathcal{T}_{Y_{0}}^{u, i n}$ to represent event 2, and $\mathcal{T}_{Y_{0}}^{u, o u t}$ to represent event 3, $\mathcal{C}_{Z_{i}^{u}, Y_{0}}^{u}$ represent event 4, and use $\mathcal{R}_{Z_{i}^{u}, Y_{0}}^{u}$ to represent event 5. Based on the above events, the end-to-end successful transmission probability of a typical MTCD can be expressed as
\begin{equation}
\begin{split}
&\operatorname{Pr}\left(\mathcal{R}_{Z_{i}^{u}, Y_{0}}^{u} \cap \mathcal{C}_{Z_{i}^{u}, Y_{0}}^{u} \cap\left(\mathcal{V}_{X_{i}^{u}, Y_{0}}^{u} \cup \mathcal{T}_{Y_{0}}^{u, i n}\right)\right) \cup \\
&\left(\mathcal{R}_{Z_{i}^{u}, Y_{0}}^{u} \cap \mathcal{C}_{Z_{i}^{u}, Y_{0}}^{u} \cap\left(\mathcal{V}_{X_{i}^{u}, Y_{0}}^{u}-\mathcal{T}_{Y_{0}}^{u, o u t}\right)\right) \\
&=\operatorname{Pr}\left(\left.\mathcal{R}_{Z_{i}^{u}, Y_{0}}^{u} \cap \mathcal{C}_{Z_{i}, Y_{0}}^{u} \cap \mathcal{V}_{X_{i}^{u}, Y_{0}}^{u}\right| \mathcal{A}_{1}\right) \cdot \operatorname{Pr}\left(\mathcal{A}_{1}\right)\\
&+\operatorname{Pr}\left(\left.\mathcal{R}_{Z_{i}^{u}, Y_{0}}^{u} \cap \mathcal{C}_{Z_{i}^{u}, Y_{0}}^{u} \cap \mathcal{T}_{Y_{0}}^{u, i n}\right| \mathcal{A}_{1}\right) \cdot \operatorname{Pr}\left(\mathcal{A}_{1}\right)\\
&+\operatorname{Pr}\left(\left.\mathcal{R}_{Z_{i}^{u}, Y_{0}}^{u} \cap \mathcal{C}_{Z_{i}^{u}, Y_{0}}^{u} \cap\left(\mathcal{V}_{X_{i}^{u}, Y_{0}}^{u}-\mathcal{T}_{Y_{0}}^{u, o u t}\right)\right| \mathcal{A}_{2}\right) \cdot \operatorname{Pr}\left(\mathcal{A}_{2}\right)
\end{split}
\label{eq2}
\end{equation}

The probability of successful end-to-end transmission is converted into the sum of $P_{1}$, $P_{2}$, and $P_{3}$ according to the full probability formula
\begin{equation}
\begin{split}
&P_{1}=\operatorname{Pr}\left(\mathcal{R}_{Z_{i}^{u}, Y_{0}}^{u} |\left(\mathcal{C}_{Z_{i}^{u}, Y_{0}}^{u} \cap \mathcal{V}_{X_{i}^{u}, Y_{0}}^{u}\right), \mathcal{A}_{1}\right) \\
&\cdot \operatorname{Pr}\left(\mathcal{C}_{Z_{i}^{u}, Y_{0}}^{u} | \mathcal{V}_{X_{i}^{u}, Y_{0}}^{u}, \mathcal{A}_{1}\right) \cdot\operatorname{Pr}\left(\left.\mathcal{V}_{X_{i}^{u}, Y_{0}}^{u}\right| \mathcal{A}_{1}\right) \cdot\operatorname {Pr}\left(\mathcal{A}_{1}\right)
\end{split}
\label{eq3}
\end{equation}
%
%
\begin{equation}
\begin{split}
&P_{2}=\operatorname{Pr}\left(\mathcal{R}_{Z_{i}^{u}, Y_{0}}^{u} |\left(\mathcal{C}_{Z_{i}^{u}, Y_{0}}^{u} \cap \mathcal{T}_{Y_{0}}^{u, i n}\right), \mathcal{A}_{1}\right) \\
&\cdot \operatorname{Pr}\left(\mathcal{C}_{Z_{i}^{u}, Y_{0}}^{u} | \mathcal{T}_{Y_{0}}^{u, i n}, \mathcal{A}_{1}\right) \cdot \operatorname{Pr}\left(\left.\mathcal{T}_{Y_{0}}^{u, i n}\right| \mathcal{A}_{1}\right)
\cdot\operatorname {Pr}\left(\mathcal{A}_{1}\right)
\end{split}
\label{eq4}
\end{equation}
%
%
\begin{equation}
\begin{split}
&P_{3}=\operatorname{Pr}\left(\mathcal{R}_{Z_{i}^{u}, Y_{0}}^{u} |\left(\mathcal{C}_{Z_{i}^{u}, Y_{0}}^{u} \cap\left(\mathcal{V}_{X_{i}^{u}, Y_{0}}^{u}-\mathcal{T}_{Y_{0}}^{u, o u t}\right)\right), \mathcal{A}_{2}\right) \\
&\cdot \operatorname{Pr}\left(\mathcal{C}_{Z_{i}^{u}, Y_{0}}^{u} \cap\left(\mathcal{V}_{X_{i}^{u}, Y_{0}}^{u}-\mathcal{T}_{Y_{0}}^{u, o u t}\right), \mathcal{A}_{2}\right) \\
&\cdot \operatorname{Pr}\left(\left.\left(\mathcal{V}_{X_{i}^{u}, Y_{0}}^{u}-\mathcal{T}_{Y_{0}}^{u, o u t}\right)\right| \mathcal{A}_{2}\right)
\cdot\operatorname {Pr}\left(\mathcal{A}_{2}\right)
\end{split}
\label{eq5}
\end{equation}
where $\mathcal{A}_{1}$ indicates that the MTCD is transferred into MTCG at $Y_{0}$, and $\mathcal{A}_{2}$ indicates that the transfer is out.

According to the poisson distribution formula, $\Pr{\left(\mathcal{A}_1\right)}$ and $\Pr{\left(\mathcal{A}_2\right)}$ can be expressed as
\begin{equation}
\begin{split}
\operatorname{Pr}\left(\mathcal{A}_{1}\right)=\operatorname{Pr}\left\{k_{i} \geq k_{0}\right\}=\sum_{k=k_{0}}^{\infty} \frac{\left(\lambda_{D} S_{Y_{i}}\right)^{k}}{k !} \exp \left(-\lambda_{D} S_{Y_{i}}\right)
\end{split}
\label{eq6}
\end{equation}
%
%
\begin{equation}
\begin{split}
\operatorname{Pr}\left(\mathcal{A}_{2}\right)=\operatorname{Pr}\left\{k_{i}<k_{0}\right\}=\sum_{k=0}^{k_{0}} \frac{\left(\lambda_{D} S_{Y_{i}}\right)^{k}}{k !} \exp \left(-\lambda_{D} S_{Y_{i}}\right)
\end{split}
\label{eq7}
\end{equation}
where $k_{0}$ represents the number of MTCDs in the $Y_{0}$ region, $k_{i}$ represents the number of MTCDs in the grouping area nearest to $Y_{0}$, and $S_{Y_{i}}$ is the area of the grouping region centered on $Y_{i}$. Obviously, the transfer in $Y_{i}$ region is the transfer out of $Y_{0}$ region, so $Pr{\left(\mathcal{A}_1\right)+Pr{\left(\mathcal{A}_2\right)}=1}$.

When MTCD transfer occurs, the number of transfer devices $k_{change}$ is
\begin{equation}
\begin{split}
k_{\text {change}}=\left\lfloor\left|\frac{k_{i}-k_{0}}{2}\right|\right\rfloor
\end{split}
\label{eq8}
\end{equation}
when $k_{0}{<k}_{i}$, it means that there are $k_{change}$ MTCDs transferred from $Y_{i}$ to $Y_{0}$ region; otherwise, it means that $k_{change}$ MTCDs are transferred from $Y_{0}$ to $Y_{i}$ region.

According to $P_{1}$, $P_{2}$, $P_{3}$, the end-to-end outage probability can be expressed as
\begin{equation}
\begin{split}
\varepsilon=E \prod_{Y_{i} \in \Phi_{G}}\left(1-\left(P_{1}+P_{2}+P_{3}\right)\right)
\end{split}
\label{eq9}
\end{equation}

Assume that $\mathrm{\Phi}_{D}^{u}$ is used to represent the location set of typical MTCD transmitted on channel \emph{u}, i.e. $X_{i}^{u}\in\mathrm{\Phi}_{D}^{u}$. $\mathcal{V}_{X_{i},Y_{0}}^{u}$ is equivalent to the event that there exists no MTCG except $Y_{0}$ within a closed ball of radius $\left\|X_{i}^{u}-Y_{0}\right\|$ centered at $X_{i}$, then
\begin{equation}
\begin{split}
&\operatorname{Pr}\left(\left.\mathcal{V}_{X_{i}^{u}, Y_{0}}^{u}\right|\mathcal{A}_{1}\right)=\\
&\operatorname{Pr}\left(\left(\Phi_{G} \backslash\left\{Y_{0}\right\}\right) \cap B\left(Z_{i}^{u},\left\|Z_{i}^{u}-Y_{0}\right\|\right)=\emptyset\right) \\
&=\exp \left(-\pi \lambda_{G}\left\|Z_{i}^{u}-Y_{0}\right\|^{2}\right)
\end{split}
\label{eq10}
\end{equation}
where $Z_{i}^{u}$ indicates the MTCD numbers in the $Y_{0}$ region after the transfer. Since $X_{i}^{u}$ represents a typical MTCD, under $\mathcal{A}_{1}$ conditions $X_{i}^{u}=Z_{i}^{u}$. Then, for event $\left.\mathcal{T}_{Y_{0}}^{u, i n}\right| \mathcal{A}_{1}$ in $P_{2}$ and event $\mathcal{V}_{X_{i}^{u}, Y_{0}}^{u}-\mathcal{T}_{Y_{0}}^{u, out}$ in $P_{3}$, the probability can be expressed as
\begin{equation}
\begin{split}
\operatorname{Pr}\left(\left.\mathcal{T}_{Y_{0}}^{u, i n}\right| \mathcal{A}_{1}\right)=\frac{k_{\text {change }}}{\sum k_{j}}
\end{split}
\label{eq11}
\end{equation}
%
%
%
\begin{equation}
\begin{split}
\operatorname{Pr}\left(\left.\left(\mathcal{V}_{X_{i}^{u}, Y_{0}}^{u}-\mathcal{T}_{Y_{0}}^{u, o u t}\right)\right| \mathcal{A}_{2}\right)=\frac{k_{0}-k_{c h a n g e}}{\sum k_{j}}
\end{split}
\label{eq12}
\end{equation}
where $\sum k_{j}$ refers to the total number of MTCDs in the region.

A typical MTCD packet can be successfully captured by MTCG located at $Y_{0}$ if the SIR of the packet is greater than the threshold $\eta$, otherwise it cannot capture it. The probability of the event $\mathcal{C}_{Z_{i}^{u}, Y_{0}}^{u} | \mathcal{V}_{X_{i}^{u}, Y_{0}}^{u}$. under the condition $\mathcal{A}_{1}$ can be expressed as
\begin{equation}
\begin{split}
\operatorname{Pr}\left(\mathcal{C}_{Z_{i}^{u}, Y_{0}}^{u} | \mathcal{V}_{X_{i}^{u}, Y_{0}}^{u}, \mathcal{A}_{1}\right)=\exp \left(-\pi \frac{\lambda_{D}}{U_{1}} \eta^{\frac{2}{\alpha}}\left\|Z_{i}^{u}-Y_{0}\right\|^{2} K_{\alpha}\right)
\end{split}
\label{eq13}
\end{equation}
where $K_{\alpha}=\int_{0}^{\infty} \frac{d t}{1+t^{\frac{\alpha}{2}}}$, $\alpha$ is the path loss index. Obviously, the event $\operatorname{Pr}\left(\mathcal{C}_{Z_{i}^{u}, Y_{0}}^{u} | \mathcal{T}_{Y_{0}}^{u, i n}, \mathcal{A}_{1}\right)$ in $P_{2}$ and the event $\operatorname{Pr}\left(\mathcal{C}_{Z_{i}^{u}, Y_{0}}^{u} \cap\left(\mathcal{V}_{X_{i}^{u} Y_{0}}^{u}-\mathcal{T}_{Y_{0}}^{u, o u t}\right), \mathcal{A}_{2}\right)$ in $P_{3}$ can also be represented by \eqref{eq13}. MTCG's average probability of capturing all MTCDs, i.e. the average capture probability $p_{c,in,\mathcal{V}}$ of the event $\mathcal{C}_{Z_{i}^{u}, Y_{0}}^{u} | \mathcal{V}_{X_{i}^{u}, Y_{0}}^{u}$ under $\mathcal{A}_{1}$ without the condition $\left\|Z_{i}^{u}-Y_{0}\right\|$ can be expressed as
\begin{equation}
\begin{split}
p_{c, i n, \mathcal{V}}=\left(\frac{\lambda_{D}}{U_{1} \lambda_{G}} \eta^{\frac{2}{\alpha}} K_{\alpha}+1\right)^{-1}
\end{split}
\label{eq14}
\end{equation}

Similarly, the MTCD average capture probability $p_{c, i n, \mathcal{T}}$ of the event $\mathcal{C}_{Z_{i}^{u}, Y_{0}}^{u} | \mathcal{T}_{Y_{0}}^{u, i n}$ can be expressed as
\begin{equation}
\begin{split}
&p_{c, i n, \mathcal{T}}=\left(\frac{\lambda_{D}}{U_{1} \lambda_{G}} \eta^{\frac{2}{\alpha}} K_{\alpha}+1\right)^{-1} \\
&\cdot \exp \left(-\left(\pi \frac{\lambda_{D}}{U_{1}} \eta^{\frac{2}{\alpha}} K_{\alpha}+\pi \lambda_{G}\right)\left\|Z_{i}^{u}-Y_{0}\right\|^{2} / 4\right)
\end{split}
\label{eq15}
\end{equation}
the detailed derivation of $p_{c, i n, \mathcal{T}}$ is shown in appendix \ref{appendix_A}.

Under condition $\mathcal{A}_{2}$, the MTCD average capture probability $p_{c,out}$ of event $\mathcal{C}_{Z_{i}^{u}, Y_{0}}^{u} \cap\left(\mathcal{V}_{X_{i}^{u}, Y_{0}}^{u}-\mathcal{T}_{Y_{0}}^{u, o u t}\right)$ can be expressed as
\begin{equation}
\begin{split}
p_{c, o u t}=\left(\frac{\lambda_{D}}{U_{1} \lambda_{G}} \eta^{\frac{2}{\alpha}} K_{\alpha}+1\right)^{-1} \cdot \frac{k_{0}-k_{\text {change}}}{k_{0}}
\end{split}
\label{eq16}
\end{equation}

All data packets can be successfully relayed when the number of data packets successfully captured by MTCG is less than $U_{2}$; otherwise, only $U_{2}$ data packets can be successfully relayed randomly, and the transmission success probability in MTCG2C link can be expressed as
\begin{equation}
\begin{split}
\operatorname{Pr}\left(\mathcal{R}_{Z_{i}^{u}, Y_{0}}^{u}\right)=\left\{\begin{array}{cc}{\frac{U_{2}}{k_{c} p_{c}}} & {, \text { if } k_{c} p_{c}>U_{2}} \\ {1} & {, \text { others }}\end{array}\right.
\end{split}
\label{eq17}
\end{equation}
where $k_{c}$ represents the number of MTCDs in the region where the MTCG is located, under condition $\mathcal{A}_{1}$, $k_{c}=k_{0}+k_{change}$, under condition $\mathcal{A}_{2}$, $k_{c}=k_{0}-k_{change}$.
$p_{c}$ represents MTCG’s average capture probability for data packets, which is recorded as $p_{c,in,\mathcal{V}}$ under event $\mathcal{C}_{Z_{i}^{u}, Y_{0}}^{u} | \mathcal{V}_{X_{i}^{u}, Y_{0}}^{u}$, $p_{c,in,\mathcal{T}}$ under event $\mathcal{C}_{Z_{i}^{u}, Y_{0}}^{u} | \mathcal{T}_{Y_{0}}^{u, i n}$, and $p_{c,out}$ under event $\mathcal{C}_{Z_{i}^{u}, Y_{0}}^{u} \cap\left(\mathcal{V}_{X_{i}^{u}, Y_{0}}^{u}-\mathcal{T}_{Y_{0}}^{u, o u t}\right)$.

According to the above derivation, the expressions of $P_{1}$, $P_{2}$, and $P_{3}$ are
\begin{equation}
\begin{split}
&P_{1}=\frac{U_{2} \cdot \exp \left(-\pi \frac{\lambda_{D}}{U_{1}} \eta^{\frac{2}{\alpha}}\left\|Z_{i}^{u}-Y_{0}\right\|^{2} K_{\alpha}\right)}{\left(k_{0}+k_{\text {change }}\right) p_{c, \text { in }, \mathcal{V}}} \\
& \cdot \exp \left(-\pi \lambda_{G}\left\|Z_{i}^{u}-Y_{0}\right\|^{2}\right) \cdot \sum_{k=k_{0}}^{\infty} \frac{\left(\lambda_{D} S_{Y_{i}}\right)^{k}}{k !} \cdot \exp \left(-\lambda_{D} S_{Y_{i}}\right)
\end{split}
\label{eq18}
\end{equation}
%
%
\begin{equation}
\begin{split}
&P_{2}=\frac{U_{2} \cdot \exp \left(-\pi \frac{\lambda_{D}}{U_{1}} \eta^{\frac{2}{\alpha}}\left\|Z_{i}^{u}-Y_{0}\right\|^{2} K_{\alpha}\right)}{\left(k_{0}+k_{\text {change }}\right) p_{c, \text { in }, \mathcal{T}}} \\
& \cdot \frac{k_{\text {change}}}{\sum k_{j}} \cdot \sum_{k=k_{0}}^{\infty} \frac{\left(\lambda_{D} S_{Y_{i}}\right)^{k}}{k ! }\cdot \exp \left(-\lambda_{D} S_{Y_{i}}\right) 
\end{split}
\label{eq19}
\end{equation}
%
%
%
\begin{equation}
\begin{split}
&P_{2}=\frac{U_{2} \cdot \exp \left(-\pi \frac{\lambda_{D}}{U_{1}} \eta^{\frac{2}{\alpha}}\left\|Z_{i}^{u}-Y_{0}\right\|^{2} K_{\alpha}\right)}{\left(k_{0}-k_{\text {change }}\right) p_{c, \text { out }}} \\
& \cdot \frac{k_{0}-k_{\text {change}}}{k_{0}} \cdot \sum_{k=k_{0}}^{\infty} \frac{\left(\lambda_{D} S_{Y_{i}}\right)^{k}}{k ! }\cdot \exp \left(-\lambda_{D} S_{Y_{i}}\right) 
\end{split}
\label{eq20}
\end{equation}

Substituting \eqref{eq18} - \eqref{eq20} into \eqref{eq9} and \eqref{eq1} can get the outage probability and transmission capacity. Specific results analysis will be given in Section \ref{section4}.
%
%
%
%
%
%
%
\section{Simulation}
\label{section4}
In this section, Monte Carlo simulations were performed to verify the accuracy of the analysis results. Changes in system transmission capacity and outage probability were evaluated when using the proposed algorithm. The analysis results are very close to the results of the actual simulation, the error is due to the signal attenuation in MTCG2C link is not considered. In addition, the performance superiority of the proposed algorithm is verified by comparing with the no outage constraint and the nearest principle relay algorithm (NPRA) \cite{b2}. Unless stated otherwise, results are obtained by setting $\eta=3$dB, $\alpha=5$, $\omega_{1}=30$, $\omega_{2}=5$, $R_{1}=1800$ and $R_{2}=1800$.

%
%
%
%
%
\begin{figure}[!t]
\centerline{\includegraphics[width=\columnwidth]{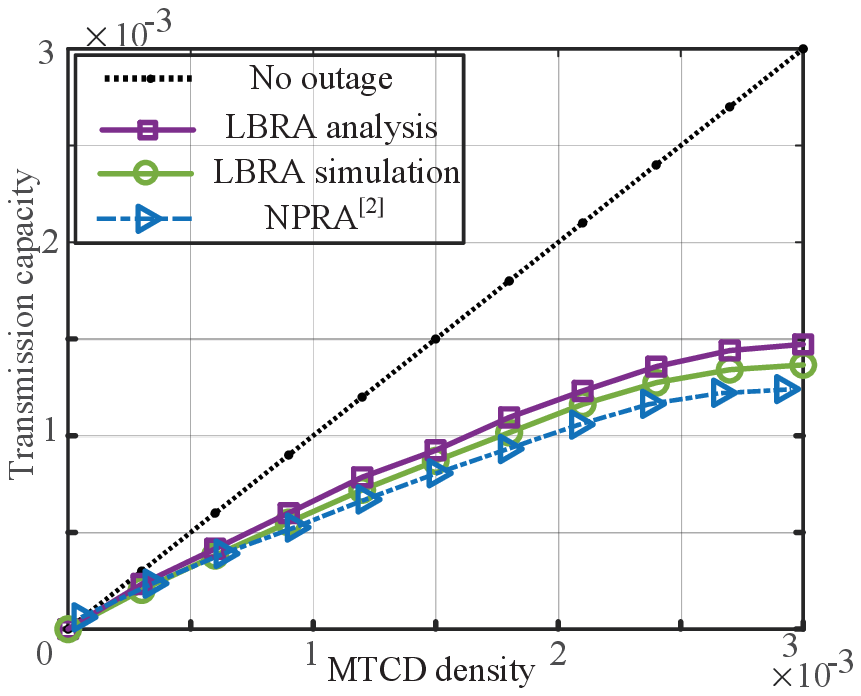}}
\caption{Relationship between MTCD density and transmission capacity, the parameter is $\lambda_{G}=1\times10^{-4}$}
\label{fig3}
\end{figure}
\begin{figure}[!t]
\centerline{\includegraphics[width=\columnwidth]{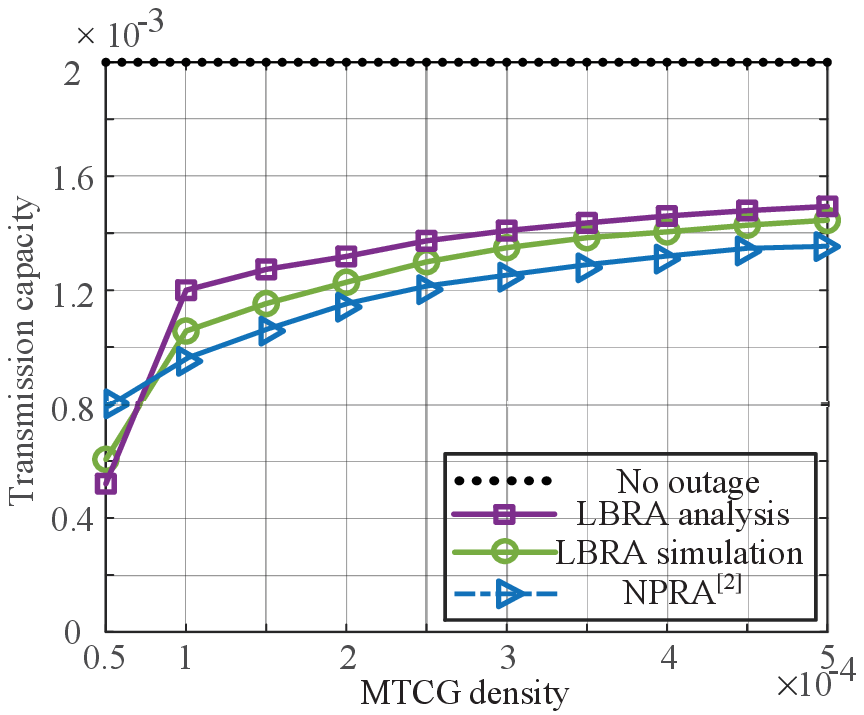}}
\caption{Relationship between MTCG density and transmission capacity, the parameter is $\lambda_{D}=2\times10^{-3}$.}
\label{fig4}
\end{figure}
\begin{figure}[!t]
\centerline{\includegraphics[width=\columnwidth]{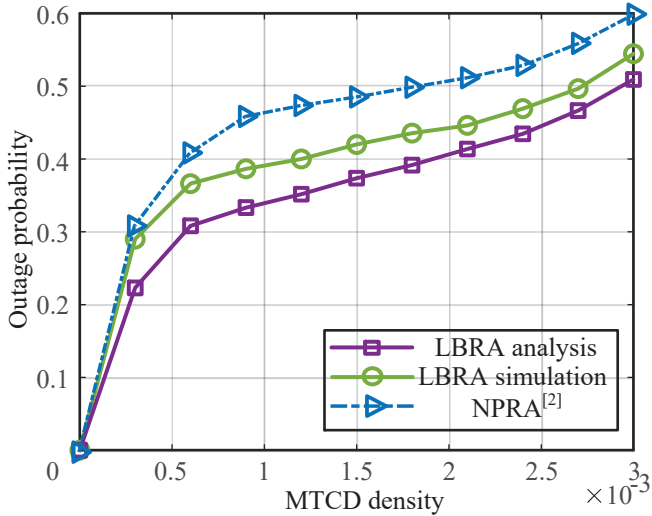}}
\caption{Relationship between MTCD density and outage probability, the parameter is $\lambda_{G}=1\times10^{-4}$.}
\label{fig5}
\end{figure}

Fig. \ref{fig3} is a trend diagram of the transmission capacity \emph{C} as a function of the MTCD density when three schemes are used. The transmission capacity increases linearly within a certain range, and at last it tends to be smooth, with the density of MTCD increases. This is due to the outage probability will gradually increase with the MTCD density increases, i.e. the increase in outage probability balances the increase in transmission capacity benefit from the increase in the density of MTCD. The increase in outage probability can be observed by comparing the difference between the performance curve and the black no outage dotted line. Finally, the transmission capacity will be stable at a fixed value, for the transmission capacity is also limited by the spectrum resources allocated by the DAC to the MTC. The performance of LBRA is better than traditional NPRA when the MTCG density $\lambda_{G}$ is fixed by observing Fig. \ref{fig3}. At $\lambda_{D}=3\times10^{-3}$, LBRA has a capacity improvement of about 0.7dB compared with traditional NPRA.

Fig. \ref{fig4} shows the trend of transmission capacity \emph{C} changing with MTCG density when three schemes are used. When the value of $\lambda_{G}$ is low, the transmission performance of LBRA is worse than NPRA, which is due to the sparse distribution of MTCG, resulting in an increase in the distance between adjacent groups. Therefore, when the MTCD density is fixed, MTCD after the transfer needs to be relayed by MTCG farther away, which leads to a significant increase in the outage probability of MTCD2G link and a corresponding decrease in its transmission capacity. However, LBRA's performance exceeded NPRA when MTCG density $\lambda_{G}$ reached near $0.75\times10^{-4}$, which is due to the distance between groups decreased, and MTCD transfer became easier with the increase of $\lambda_{G}$, making reasonable use of MTCG2C link spectrum resources. Subsequently, LBRA has consistently outperformed than NPRA in performance with high MTCG density. LBRA had a capacity increase of about 0.46dB compared to the traditional NPRA at $\lambda_{G}=5\times10^{-4}$.

Fig. \ref{fig5} shows the relationship between outage probability $\varepsilon$ and MTCD density $\lambda_{D}$ when two algorithms are used. The outage probability curve in the figure shows a significant inflection point near $\lambda_{D}=3\times10^{-3}$, which is due to the derivation of the outage probability of MTCG2C link results in a piecewise function based on theoretical analysis. In addition, the outage probability began to increase rapidly near the $\lambda_{D}=2.75\times10^{-3}$, as the number of packets captured by MTCG began to exceed the number that could be relayed to the DAC. The LBRA always has the lowest outage probability at any MTCD density by comparing the two algorithms. Compared with the traditional NPRA algorithm, the LBRA algorithm has approximately 0.8dB outage probability improvement at $\lambda_{D}=3\times10^{-3}$.

The above simulation results show that compared with the traditional resource allocation algorithm NPRA, the LBRA proposed in this paper can reduce the outage probability and increase the transmission capacity while maintaining a high MTCD connection density. It is of great significance in practical applications. As in the smart port scenario, the use of LBRA can support more unmanned forklift work, ensure a high probability of successful connection between the forklift and the DAC, and improve the efficiency of cargo transportation.
%
%
%
%
%
%
%
\section{Conclusion}
In this paper, a load balancing algorithm is proposed which reallocate spectrum resources on MTCD2G link to solve the problem of unfair resource allocation in traditional mobile MTC relay. Numerical results show that LBRA has good performance, especially when MTCD density is high, its transmission capacity and outage probability performance are better than the traditional algorithm, indicating that the proposed algorithm is suitable for MTCD intensive deployment environment. Nonetheless, there are some improvements that can be made in this paper: pilot allocation may also need to be considered by the DAC; MTCD grouping can also consider other methods such as business type; MTCG may not be able to communicate with both DAC and MTCD at the same time due to the limitations of its own communication mechanism in practice.

\appendices
\section{}
\label{appendix_A}
For the calculation of the average capture probability of packets transferred from $Y_{i}$, the difference lies in the integration interval. The integral for a typical package starts at 0, and here the integral starts at $\left\| {{Y}_{0}}-{{Y}_{i}} \right\|/2$.
\begin{equation}
\begin{aligned}
  & {{p}_{c,in,\mathsf{\mathcal{T}}}}=\int_{\left\| {{Y}_{0}}-{{Y}_{i}} \right\|/2}^{\infty }{\exp \left( -\pi \frac{\lambda {}_{D}}{{{U}_{1}}}{{\eta }^{\frac{2}{\alpha }}}{{\left\| Z_{i}^{u}-{{Y}_{0}} \right\|}^{2}}{{K}_{\alpha }} \right)} \\ 
 & \cdot 2\pi r{{\lambda }_{G}}\exp \left( -{{\lambda }_{G}}\pi {{r}^{2}} \right)dr \\ 
 & =2\pi {{\lambda }_{G}}\int_{\left\| {{Y}_{0}}-{{Y}_{i}} \right\|/2}^{\infty }{\exp \left( -\pi \left( \frac{\lambda {}_{D}}{{{U}_{1}}}{{\eta }^{\frac{2}{\alpha }}}{{K}_{\alpha }}+{{\lambda }_{G}} \right){{r}^{2}} \right)}rdr \\ 
\end{aligned}
\label{eq21}
\end{equation}

Let's $W=\frac{\lambda {}_{D}}{{{U}_{1}}}{{\eta }^{\frac{2}{\alpha }}}{{K}_{\alpha }}+{{\lambda }_{G}}$, and substitute it into (\ref{eq21})
\begin{equation}
\begin{aligned}
& {{p}_{c,in,\mathsf{\mathcal{T}}}}=2\pi {{\lambda }_{G}}\int_{\left\| {{Y}_{0}}-{{Y}_{i}} \right\|/2}^{\infty }{\exp \left( -\pi W{{r}^{2}} \right)}rdr \\ 
 & =\frac{{{\lambda }_{G}}}{W}\int_{\left\| {{Y}_{0}}-{{Y}_{i}} \right\|/2}^{\infty }{\exp \left( -\pi W{{r}^{2}} \right)}d\pi W{{r}^{2}} \\ 
 & ={{\left( \frac{\lambda {}_{D}}{{{U}_{1}}{{\lambda }_{G}}}{{\eta }^{\frac{2}{\alpha }}}{{K}_{\alpha }}+1 \right)}^{-1}} \\ 
 & \cdot \exp \left( -\left( \pi \frac{\lambda {}_{D}}{{{U}_{1}}}{{\eta }^{\frac{2}{\alpha }}}{{K}_{\alpha }}+\pi {{\lambda }_{G}} \right){{\left\| {{Y}_{0}}-{{Y}_{i}} \right\|}^{2}}/4 \right) \\ 
\end{aligned}
\label{eq22}
\end{equation}


\end{document}